\documentclass[preprint,showpacs,preprintnumbers,amsmath,amssymb]{revtex4}

\usepackage{graphicx}
\usepackage{dcolumn}
\usepackage{bm}

\begin{document}


\title{A novel method to construct stationary solutions of the Vlasov-Maxwell system}

\author{Akihiro Suzuki}
\email{suzuki@resceu.s.u-tokyo.ac.jp}
\altaffiliation[Also at ]{Department of Astronomy, Graduate School of Science, University of Tokyo, Bunkyo-ku, Tokyo 113-0033, Japan}
\author{Toshikazu Shigeyama}%
\affiliation{Research center for the Early Universe, Graduate School of Science, University of Tokyo, Bunkyo-ku, Tokyo 113-0033, Japan
}%

\date{\today}

\begin{abstract}
A novel method to derive stationary solutions of the Vlasov-Maxwell system is established. 
This method is based on the assumption that the deviation of the velocity distribution from the Maxwell-Boltzmann distribution can be expanded by the Hermite polynomials. 
By applying our method, a new two-dimensional equilibrium is derived, which may provide an initial setup for investigations of three-dimensional collisionless reconnection of magnetic fields. 
\end{abstract}

\pacs{Valid PACS appear here}
\maketitle

\section{\label{sec:level1}INTRODUCTION}
It goes without saying that magnetized plasmas play important roles both in nature and laboratory experiments. 
 For example, in the solar atmosphere and the terrestrial magnetotail, magnetic reconnection \cite{parker,sweet,petschek} is considered to be a plausible mechanism converting magnetic energy into kinetic energy of particles. 
Furthermore, in supernovae, gamma-ray bursts, and other astrophysical jets, some plasma instabilities develop when the ejecta interact with the ambient medium \cite{ML}. 
Such plasma instabilities are important also in laser experiments. 

There are some approaches to describe the behavior of such plasmas and electromagnetic fields. 
One is the ideal or non-ideal magnetohydrodynamics (MHD). 
This is based on the fluid treatment implicitly assuming the Maxwellian velocity distribution for particles. 
On the other hand, the kinetic treatment of plasmas leads to the Boltzmann equation coupled with the Maxwell equations (see, for example, Ref. \cite{CK02}). 
By using this approach, we can deal with plasmas even if the velocity distribution deviates from the Maxwellian. 
Neglecting the effect of collisions, we obtain the Vlasov-Maxwell system that can describe sufficiently rarefied plasmas. 

In particular, the magnetic reconnection is of great interest as an active phenomenon of plasmas. 
This phenomenon, which also occurs in collisionless plasmas of the solar atmosphere and the terrestrial magnetotail, requires an anomalous resistivity, inertia-based processes or off-diagonal terms of the electron pressure tensor. 
Recently, a great deal of investigations of the collisionless magnetic reconnection have been performed by theoretical analyses \cite{LPV,CLP,DL,Galeev,LP}, particle simulations \cite{KK,PCPK91,HS97,SSJTZ03}, MHD simulations \cite{OP93,BSD95}, and a Vlasov simulation \cite{SG06}. 

Stationary solutions of the Vlasov-Maxwell system provide self-consistent configurations of collisionless plasmas and they are the convenient initial setup for investigations of the collisionless magnetic reconnection. 
Therefore, some authors have proposed such solutions \cite{harris,bennett,BGK57,mahajana,mahajanb,AP99,BBSS01,mottez,CMPC05,MP07}. 
The Harris sheet \cite{harris} is one of the most well-known equilibrium assuming the Maxwellian velocity distribution. 
This equilibrium is widely used as an initial setup for simulations of the magnetic reconnection. 
The Bennett pinch \cite{bennett} also assumes the Maxwellian velocity distribution and provides an equilibrium in cylindrical configuration. 
The so-called BGK solution \cite{BGK57} treats an electrostatic equilibrium. 
The effect of the temperature gradient is included in Ref. \cite{mahajana,mahajanb,MP07}.
Ref. \cite{CMPC05} investigates two-dimensional case.

In the present paper, we focus on a method to construct stationary solutions of the Vlasov-Maxwell system. 
By applying our method, we can reproduce some results of previous studies. 
In other words, our method is a generalization of the methods presented in these studies: 
it can treat two-dimensional plasmas with the velocity distribution deviating from the Maxwellian. 
The generalization enables us to explore new solutions. 
By way of example, we construct a new two-dimensional equilibrium, which may provide an initial setup for three-dimensional magnetic reconnection studies \cite{pritchett,SDRD2001,zeiler}. 

This paper is organized as follows. 
In Sec.\ref{form}, we describe the procedure of our method. 
In Sec.\ref{maxwell}, some results of previous studies are reproduced. 
We propose a new equilibrium derived by applying our method in Sec.\ref{new}. 
We conclude this paper in Sec.\ref{conc}. 
Appendix gives some algebraic relations used in this paper.

\section{\label{form}FORMULATION}
We deal with the Vlasov-Maxwell system.
The Vlasov equation describes the kinetic evolution of the distribution function $f_j(t,\mathbf{x},\mathbf{v})$ of particles $j$ ($=i$ for ions and $e$ for electrons) in the phase space $(x,y,z,v_x,v_y,v_z)$ and the Maxwell equations describe the evolution of the electromagnetic fields. 
Here the cartesian coordinates in the real space are $(x,y,z)$ and the corresponding coordinates in the velocity space are $(v_x,v_y,v_z)$. 
This system provides an exact model for the behavior of collisionless plasmas (see, for example, Ref. \cite{sturrock}). 
In this section, we derive stationary solutions of the Vlasov-Maxwell system under the assumptions described in the following subsection. 
\subsection{Equations}
We consider a stationary plasma. 
It is assumed to be homogeneous in the z-direction. 
Thus physical variables do not depend on $t$ and $z$. 
In this circumstances, the Vlasov equation expressed as
\begin{equation}
v_x\frac{\partial f_j}{\partial x}+
v_y\frac{\partial f_j}{\partial y}+
\frac{q_j}{m_jc}\left(cE_x-v_zB_y\right)\frac{\partial f_j}{\partial v_x}
+\frac{q_j}{m_jc}\left(cE_y+v_zB_x\right)\frac{\partial f_j}{\partial v_y}
+\frac{q_j}{m_jc}\left(v_xB_y-v_yB_x\right)\frac{\partial f_j}{\partial v_z}
=0, 
\label{vlasov}
\end{equation}
governs the distribution function $f_j$ for particles $j$ with the charge $q_j$ and the mass $m_j$. 
Here $c$ is the speed of light. $E_x,E_y$ and $B_x,B_y$ represent the $x,y$-components of electric and magnetic fields, which are functions of $x$ and $y$.
These electromagnetic fields are written as
\begin{subequations}
  \begin{eqnarray}
    E_x=-\frac{\partial \phi}{\partial x}&,&E_y=-\frac{\partial \phi}{\partial y},\\
    B_x=\frac{\partial A_z}{\partial y}&,&B_y=-\frac{\partial A_z}{\partial x},
  \end{eqnarray}
  \label{pot}
\end{subequations}
by introducing the scalar potential $\phi(x,y)$ and the $z$-component of the vector potential $A_z(x,y)$. 
The other components are assumed to vanish. 
These potentials satisfy the Poisson equations;
\begin{subequations}
  \begin{eqnarray}
    \frac{\partial^2\phi}{\partial x^2}+\frac{\partial^2\phi}{\partial y^2}&=& 
    -4\pi\rho,\\
    \frac{\partial^2A_z}{\partial x^2}+\frac{\partial^2A_z}{\partial y^2}&=& 
    -4\pi j_z,
  \end{eqnarray}
  \label{poisson}
\end{subequations}
where $\rho(x,y)$ and $j_z(x,y)$ are the charge density and the $z$-component of the electric current density, respectively. These are expressed in terms of $f_j(x,y,v_x,v_y,v_z)$ as
\begin{subequations}
  \begin{eqnarray}
    \rho&=&\sum_jq_j\int^\infty_{-\infty}dv_x\int^\infty_{-\infty}dv_y\int^\infty_{-\infty}dv_zf_j,\\
    j_z&=&\sum_j\frac{q_j}{c}\int^\infty_{-\infty}dv_x\int^\infty_{-\infty}dv_y\int^\infty_{-\infty}dv_zv_zf_j,
  \end{eqnarray}
  \label{current}
\end{subequations}
which close the system. 
Solutions of Equations (\ref{vlasov})-(\ref{current}) give self-consistent configurations of collisionless plasmas.

\subsection{Derivation of stationary solutions}
In seeking solutions of Equations (\ref{vlasov})-(\ref{current}), we assume that the distribution function takes the following form;
\begin{equation}
  f_j=\left[\sum_{n=0}^\infty g_{j,n}(A_z)H_n\left(v_z/v_j\right)\right]
  \frac{n_j}{\pi^{3/2}v_j^3}\exp\left(-\frac{v_x^2+v_y^2+v_z^2}{v_j^2}
    -\frac{2q_j\phi}{m_jv_j^2}\right),
  \label{fj}
\end{equation}
where $n_j$ and $v_j$ are the average number density and the thermal velocity of particles $j$, respectively. 
We assume they are constant. 
$H_n$ represents the $n$th-order Hermite polynomial and $g_{j,n}$ is its coefficient, which is assumed to be a function of $A_z$.
The distribution function in this form describes the deviation from the Maxwellian in terms of the Hermite polynomial expansion. 
The spatial dependence of the distribution function is described through those of potentials $\phi$ and $A_z$. 

At first, we substitute the expressions for fields (\ref{pot}) and the distribution function (\ref{fj}) into Equation (\ref{vlasov}). 
Using the recurrence relation (\ref{diff}), we obtain
\begin{equation}
\left\{\sum_{n=0}^\infty \left[\frac{dg_{j,n}}{dA_z}-\frac{2(n+1)q_j}{m_jv_jc}g_{j,n+1}\right]H_n(v_z/v_j)\right\}
\frac{n_j}{\pi^{3/2}v_j^3}\exp\left(-\frac{v_x^2+v_y^2+v_z^2}{v_j^2}
-\frac{2q_j\phi}{m_jv_j^2}\right)=0.
\end{equation}
Then we multiply both sides of this equation by $H_m(v_z/v_j)$ and integrate with respect to $v_z$ to obtain the recurrence formula for $g_{j,n}$, 
\begin{equation}
g_{j,n+1}=\frac{m_jv_jc}{2(n+1)q_j}\frac{dg_{j,n}}{dA_z}.
\label{rel}
\end{equation}
Here we have used the orthogonality relation of the Hermite polynomials (\ref{auth}). 
As a result, $g_{j,n}(n\geq 1)$ are determined once $g_{j,0}$ is given as a function of $A_z$. 

On the other hand, substitution of Equations (\ref{fj}) and (\ref{rel}) into Equations (\ref{current}) leads to
\begin{subequations}
  \begin{eqnarray}
    \rho&=&n_iq_ig_{i,0}\exp\left(-\frac{2q_i\phi}{m_iv_i^2}\right)
      +n_eq_eg_{e,0}\exp\left(-\frac{2q_e\phi}{m_ev_e^2}\right),\\
    j_z&=&\frac{m_in_iv_i^2}{2}\frac{dg_{i,0}}{dA_z}\exp\left(-\frac{2q_i\phi}{m_iv_i^2}\right)+\frac{m_en_ev_e^2}{2}\frac{dg_{e,0}}{dA_z}\exp\left(-\frac{2q_e\phi}{m_ev_e^2}\right),
  \end{eqnarray}
\label{rhoj}
\end{subequations}
where the orthogonality relation (\ref{auth}) has been used again. 
Substituting these expressions into Equations (\ref{poisson}), we obtain 
\begin{subequations}
  \begin{eqnarray}
    \frac{\partial^2\phi}{\partial x^2}+\frac{\partial^2\phi}{\partial y^2}&=& 
    -4\pi\left[n_iq_ig_{i,0}\exp\left(-\frac{2q_i\phi}{m_iv_i^2}\right)
      +n_eq_eg_{e,0}\exp\left(-\frac{2q_e\phi}{m_ev_e^2}\right)\right],
    \label{poisson2a}\\
    \frac{\partial^2A_z}{\partial x^2}+\frac{\partial^2A_z}{\partial y^2}&=& 
    -2\pi\left[\frac{m_in_iv_i^2}{2}\frac{dg_{i,0}}{dA_z}\exp\left(-\frac{2q_i\phi}{m_iv_i^2}\right)+\frac{m_en_ev_e^2}{2}\frac{dg_{e,0}}{dA_z}\exp\left(-\frac{2q_e\phi}{m_ev_e^2}\right)\right].
    \label{poisson2b}
  \end{eqnarray}
  \label{poisson2}
\end{subequations}
Thus, once we determine the potentials $\phi$ and $A_z$ as a function of $x$ and $y$ under a certain boundary condition, we obtain the distributions $g_{j,0}$ for each species, and \textit{vice versa}. 

Here we note restriction of the method described above. 
In general, a stationary equilibrium can have various velocity distributions as long as the pressure balance is achieved. 
However, Equations (\ref{poisson2}) look as if there is a one-to-one correspondence between the field configuration and the distributions $g_{j,0}$. 
This disagreement comes from the assumption that the deviation of the velocity distribution from the Maxwellian can be expanded by the Hermite polynomials. 
Therefore, among a lot of possible equilibria, the stationary solutions derived above cover a part of them. 

\section{\label{maxwell}Maxwellian equilibrium}
In this section, we consider applications of the stationary solutions derived above. 
Most of familiar stationary solutions to the Vlasov-Maxwell system are based on the classical Maxwell-Boltzmann distribution. 
For example, the Harris sheet equilibrium \cite{harris} and the Bennet pinch equilibrium \cite{bennett} are well-known one-dimensional equilibria. 
Our solutions include these equilibria. 
We reproduce the Harris sheet equilibrium in \ref{ha} and the Bennett pinch equilibrium in \ref{be} by using our method.

At first, we assume that $g_{j,0}$ is expressed as
\begin{equation}
  g_{j,0}=\exp\left(\frac{2q_ju_jA_z}{m_jv_j^2c}\right),
  \label{g0}
\end{equation}
where $u_j$ is a constant with the dimension of velocity. 
Then, using relation (\ref{rel}), we obtain
\begin{equation}
  g_{j,n}=\frac{1}{n!}\left(\frac{u_j}{v_j}\right)^n
  \exp\left(\frac{2q_ju_jA_z}{m_jv_j^2c}\right).
\end{equation}
Substitution of this expression into Equation (\ref{fj}) yields
\begin{equation}
  f_j=\left[\sum_{n=0}^\infty \frac{1}{n!}\left(\frac{u_j}{v_j}\right)^n
    H_n\left(v_z/v_j\right)\right]
  \frac{n_j}{\pi^{3/2}v_j^3}\exp\left[-\frac{v_x^2+v_y^2+v_z^2}{v_j^2}
    -\frac{2q_j}{m_jv_j^2}\left(\phi-\frac{u_jA_z}{c}\right)\right].
\end{equation}
By performing the summation with respect to $n$ in this equation through the relation (\ref{gen}), we obtain a shifted Maxwell-Boltzmann distribution;
\begin{equation}
  f_j=\frac{n_j}{\pi^{3/2}v_j^3}\exp\left[-\frac{v_x^2+v_y^2+(v_z-u_j)^2}{v_j^2}
      -\frac{2q_j}{m_jv_j^2}\left(\phi-\frac{u_jA_z}{c}\right)\right].
\end{equation}
From this expression, it is found that $u_j$ is interpreted as the bulk velocity of species $j$. 

On the other hand, substitution of Equation (\ref{g0}) into Equation (\ref{poisson2}) leads to
\begin{subequations}
  \begin{eqnarray}
    \frac{\partial^2\phi}{\partial x^2}+\frac{\partial^2\phi}{\partial y^2}&=& 
    -4\pi\left\{n_iq_i\exp\left[-\frac{2q_i}{m_iv_i^2}\left(\phi-\frac{u_iA_z}{c}\right)\right]
    \right.\\\nonumber&&\hspace{5em}\left.
    +n_eq_e\exp\left[-\frac{2q_e}{m_ev_e^2}\left(\phi-\frac{u_eA_z}{c}\right)\right]\right\},
    \label{maxa}\\
    \frac{\partial^2A_z}{\partial x^2}+\frac{\partial^2A_z}{\partial y^2}&=& 
    -\frac{4\pi}{c}\left\{n_iq_iu_i\exp\left[-\frac{2q_i}{m_iv_i^2}\left(\phi-\frac{u_iA_z}{c}\right)\right]
    \right.\\\nonumber&&\hspace{5em}\left.
    +n_eq_eu_e\exp\left[-\frac{2q_e}{m_ev_e^2}\left(\phi-\frac{u_eA_z}{c}\right)\right]\right\}.\label{maxb}
  \end{eqnarray}
  \label{max}
\end{subequations}
In order to obtain the electromagnetic fields as a function of $x$ and $y$, we need to solve these equations under a certain boundary condition. 
Although Equations (\ref{max}) seem to be complicated, we can derive some analytic solutions under appropriate assumptions. 

In the following derivation, we assume that the plasma comprises of electrons and ions with same charge but the opposite sign ($q_i=-q_e=e$) and satisfies the charge neutrality ($n_i=n_e=n_0$) and that the electric field strength is sufficiently small ($\phi=0$). 
Then, Equation (\ref{maxa}) leads to
\begin{equation}
\frac{u_i}{m_iv_i^2}=-\frac{u_e}{m_ev_e^2},
\label{vel}
\end{equation}
which implies that electrons and ions move in opposite directions. 

\subsection{\label{ha}Harris sheet equilibrium}
The Harris sheet equilibrium \cite{harris} is a well-known Vlasov-Maxwell equilibrium, which is widely used in various analytical and numerical studies of collisionless plasmas. 
For example, this equilibrium provides the initial setup for investigations of the collisionless tearing mode instabilities. 

In order to derive the Harris sheet, 
we assume that the plasma is homogeneous in $y$-direction.
Combining with the relation (\ref{vel}), Equation (\ref{maxb}) reduces to the following second-order ordinary differential equation;
\begin{equation}
  \frac{d^2A_z}{dx^2}=
  -\frac{4\pi n_0e(u_i-u_e)}{c}\exp\left(\frac{2eu_iA_z}{m_iv_i^2c}\right).
\end{equation}
This equation has a solution in the form of
\begin{equation}
A_z=-B_0L\ln[\cosh(x/L)],
\label{A_z}
\end{equation}
where $B_0$ and $L$ are constants satisfying the following relations;
\begin{equation}
B_0^2=4\pi n_0(m_iv_i^2+m_ev_e^2),
\label{b02}
\end{equation}
and
\begin{equation}
B_0L=\frac{m_iv_i^2c}{eu_i}=-\frac{m_ev_e^2c}{eu_e}.
\label{b0L}
\end{equation}
The expression (\ref{A_z}) gives a sheared magnetic field;
\begin{equation}
B_y=B_0\tanh(x/L),
\end{equation}
and a distribution function in the form of
\begin{equation}
f_j\propto \frac{1}{\cosh^2(x/L)}.
\end{equation}
From these expressions, $B_0$ is interpreted as the asymptotic field in the limit of $x\rightarrow\infty$ and $L$ the scale-length. 
\subsection{\label{be}Bennett pinch equilibrium}
The Bennett pinch \cite{bennett} is a Vlasov-Maxwell equilibrium in the cylindrical configuration. 
In cylindrical symmetry, Equation (\ref{maxb}) reduces to the following second-order ordinary differential equation.
\begin{equation}
  \frac{1}{r}\frac{d}{dr}\left(r\frac{dA_z}{dr}\right)=
  -\frac{4\pi n_0e(u_i-u_e)}{c}\exp\left(\frac{2eu_iA_z}{m_iv_i^2c}\right),
\end{equation}
where $r=\sqrt{x^2+y^2}$. 
This equation has a solution in the form of
\begin{equation}
A_z=-B_0L\ln\left[1+r^2/(4L^2)\right].
\end{equation}
Also in this case, two constants satisfy the relations (\ref{b02}) and (\ref{b0L}). 
Then we obtain a self-consistent cylindrical configuration of the plasma;
\begin{equation}
f_j\propto \frac{1}{\left[1+r^2/(4L^2)\right]^2},
\end{equation}
and it generates a magnetic field surrounding itself;
\begin{equation}
B_x=-\frac{2B_0Ly}{r^2+4L^2},\ \ \ 
B_y=\frac{2B_0Lx}{r^2+4L^2}.
\end{equation}
In this case, $B_0$ and $L$ are interpreted as the scale of the magnetic field strength and the scale of the current tube, respectively.

\section{\label{new}Non-Maxwellian equilibrium}
In the previous section, we have dealt with one-dimensional Vlasov-Maxwell equilibria based on a shifted Maxwell-Boltzmann distribution. 
Furthermore, by applying our method, we can construct two-dimensional equilibria with a distribution function deviating from the Maxwell-Boltzmann distribution as follows. 

First, we assume the charge neutrality and a small electric field strength again.
Then, Equation (\ref{poisson2a}) implies that ions and electrons have the same distribution;
\begin{equation}
g_{i,0}=g_{e,0}.
\label{same}
\end{equation}
Furthermore, we assume that $A_z$ takes the following form, 
\begin{equation}
  A_z=-B_0L[\cos(x/L)+\cos(y/L)],
  \label{vector}
\end{equation}
which provides a sinusoidal magnetic field as
\begin{equation}
  B_x=B_0\sin(y/L),\ \ \ 
  B_y=-B_0\sin(x/L).
\end{equation}
Substituting Equations (\ref{same}) and (\ref{vector}) into Equation (\ref{poisson2b}), we obtain the following first-order ordinary differential equation;
\begin{equation}
\frac{dg_{i,0}}{dA_z}=\frac{A_z}{2\pi n_0L^2(m_iv_i^2+m_ev_e^2)}.
\end{equation}
Integration of this equation gives
\begin{equation}
g_{i,0}=g_{e,0}=\frac{A_z^2+C}{4\pi n_0L^2(m_iv_i^2+m_ev_e^2)},
\end{equation}
where $C$ is a constant of integration. 
Substitution of this expression into Equation (\ref{rel}) leads to
\begin{eqnarray}
g_{j,1}&=&\frac{m_jv_jcA_z}{4\pi q_jn_0L^2(m_iv_i^2+m_ev_e^2)},\\
g_{j,2}&=&\frac{m_j^2v_j^2c^2}{16\pi q_j^2n_0L^2(m_iv_i^2+m_ev_e^2)},\\
g_{j,n}&=&0\ \ \ \mathrm{for}\ n\geq 3.
\end{eqnarray}
As a result, we obtain the following distribution function after simple algebraic manipulations;
\begin{equation}
f_j=\frac{1}{4\pi^{5/2}L^2(m_iv_i^2+m_ev_e^2)v_j^3}
\left[\left(A_z+\frac{m_jc}{q_j}v_z\right)^2+A_0^2\right]
\exp\left(-\frac{v_x^2+v_y^2+v_z^2}{v_j^2}\right).
\label{fjnew}
\end{equation}
Here, the explicit expressions for $H_0$, $H_1$, and $H_2$ (\ref{hermite}) have been used and  
$A_0$ is a constant defined as 
\begin{equation}
A_0^2=C-\frac{m_jv_j^2c}{2q_j^2},
\label{C}
\end{equation}
in order that the distribution function (\ref{fjnew}) takes positive values at arbitrary points in the phase space. 
This distribution satisfies the Vlasov-Maxwell system (\ref{vlasov}) - (\ref{current}) exactly. 

\begin{figure}
\includegraphics{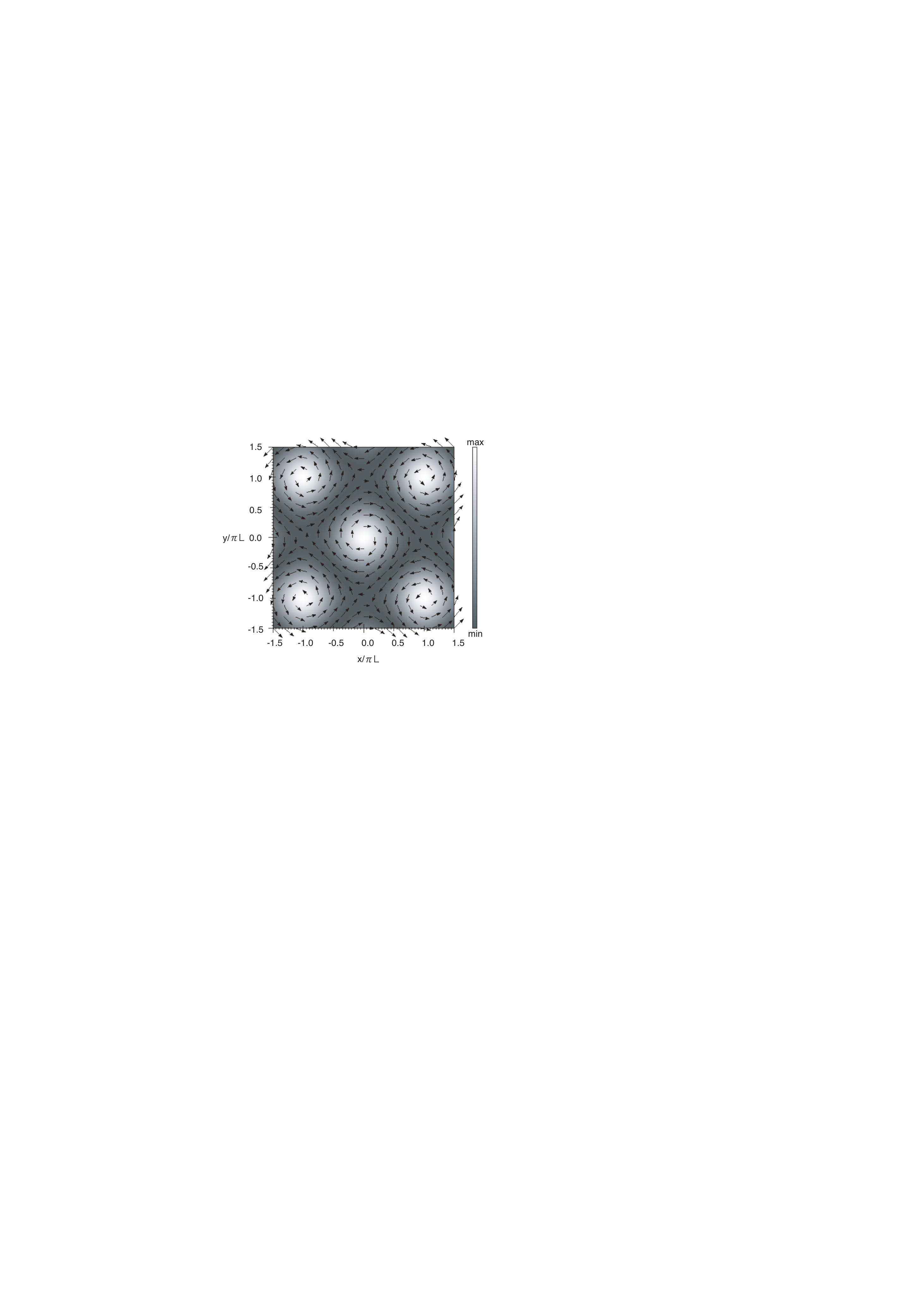}
\caption{The configuration of the two-dimensional Vlasov-Maxwell equilibrium which is derived by applying our method. 
The gray scale represents the density distribution of electrons (or ions) in arbitrary units and the arrows represent the magnetic fields.}
\label{figure1}
\end{figure}

Figure \ref{figure1} illustrates the thus derived configuration of equilibrium.
The gray scale and the arrows represent the density distribution $g_{j,0}$ of electrons or ions in arbitrary units and the magnetic field, respectively. 
We can see that the current filaments lie along the $z$-axis and generate the magnetic fields around themselves. 
Each filament is surrounded by four filaments that carry anti-parallel currents. 

\begin{figure}
\includegraphics{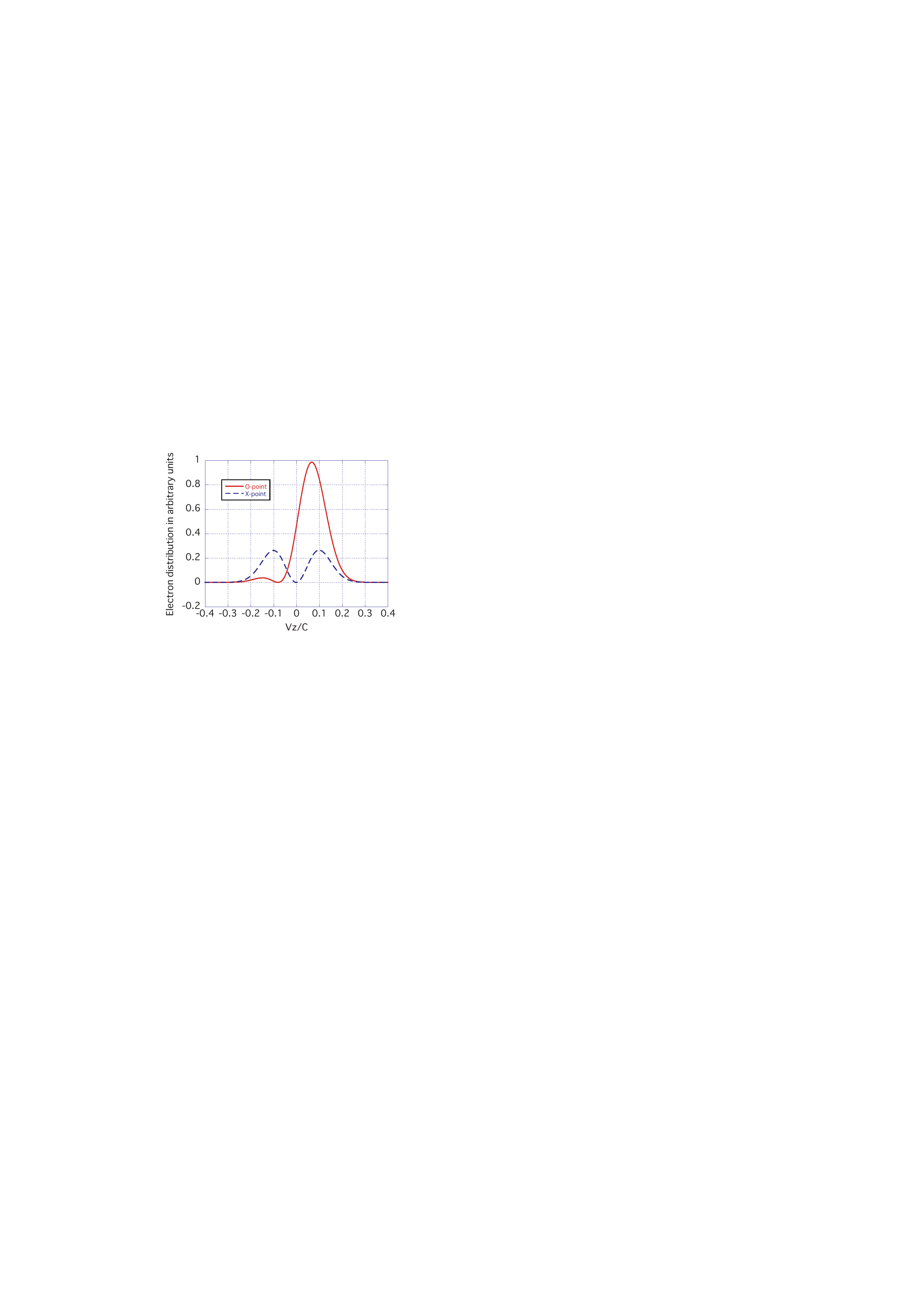}
\caption{(Color online) The velocity distributions of electrons at the O-point (solid line) and the X-point (dashed line).}
\label{figure2}
\end{figure}

Here, we focus on the velocity distributions at two points. 
One is the center of a filament (refer to as the O-point) and the other is a point between the filaments (refer to as the X-point). 
The magnetic field at the X-point is sheared one like the Harris sheet equilibrium. 
Figure \ref{figure2} shows the velocity distributions of electrons at the X and O-point, respectively. 
The values of parameters are as follows; 
$B_0=0.04m_e\omega_e/e$, where $\omega_e$ is the electron plasma frequency, $A_0=0$, and $v_e=0.1c$. 
At the X-point, we see the double-peak distribution. 
On the other hand, a distribution similar to the shifted Maxwellian one is achieved at the O-point as long as $eB_0L>m_ecv_e$ holds. 

Finally, we mention the velocity distribution of ions. 
Because of the large mass ($m_i\simeq 2000m_e$) and the small thermal velocity ($v_i\ll v_e$) according to Equation (\ref{C}), 
the distribution (\ref{fjnew}) has a sharp double-peak like two delta functions and becomes almost independent of $x$ and $y$.   

\section{\label{conc}CONCLUSIONS}
In this paper, we have developed a novel method to construct stationary solutions of the Vlasov-Maxwell system. 
This investigation must lead to great insight into the properties of self-consistent plasma configurations. 
By applying our method, some of previous studies are reproduced and a new two-dimensional equilibrium is proposed. 
In particular, it is found that a sheared magnetic field like that in the Harris sheet equilibrium is generated. 
Although the stability of the equilibrium configuration needs to be investigated, it may provide an initial setup for investigations of the collisionless magnetic reconnection that takes into account three-dimensional effects. 

\begin{acknowledgments}
This work is supported in part by Grant-in-Aid for Scientific Research (16540213) from the Ministry of Education, Culture, Sports, Science, and Technology of Japan and JSPS (Japan Society for Promotion of Science) Core-to-Core Program ``International Research Network for
Dark Energy''.
\end{acknowledgments}

\appendix
\section{PROPERTIES OF HERMITE POLYNOMIALS}
In this section, we review the properties of the Hermite polynomials used in this paper (see, for example, \cite{abramowitz}).
The $n$th-order Hermite polynomial $H_n(x)$ is defined as the solutions of the Hermite's differential equation;
\begin{equation}
\frac{d^2H_n}{dx^2}-2x\frac{dH_n}{dx}+2nH_n=0,
\end{equation}
The set $H_n(x)$ forms an orthogonal basis in the function space.
In other words, the Hermite polynomials satisfy the orthogonality relation;
\begin{equation}
\int^\infty_{-\infty}H_m(x)H_n(x)e^{-x^2}dx=
2^n\sqrt{\pi}n!\delta_{m,n},
\label{auth}
\end{equation}
where $\delta_{m,n}$ represents Kronecker's delta. 
The Hermite polynomials are also given by the generating function as
\begin{equation}
  e^{-t^2+2xt}=\sum_{n=0}^\infty\frac{t^n}{n!}H_n(x).
  \label{gen}
\end{equation}
From this equation, we can obtain the expression for the Hermite polynomials. 
The first three are
\begin{equation}
H_0=1,\ \ \ H_1=2x,\ \ \ H_2=4x^2-2.
\label{hermite}
\end{equation}
The derivative of $H_n$ is expressed in terms of $H_{n-1}$ as
\begin{equation}
\frac{dH_n}{dx}=2nH_{n-1}.
\label{diff}
\end{equation}

\newpage

\end{document}